\documentclass[a4paper,11pt]{article}
\usepackage{pos}

\title{Probing 95 GeV Higgs in the 2HDM Type-III}

\author[a,b]{A. Belyaev}
\author[c]{R. Benbrik}
\author[c]{M. Boukidi} 
\author[a]{M. Chakraborti}
\author[a,d]{S. Moretti}
\author*[a,b]{S. Semlali}

\vspace{0.1cm}

\affiliation[a]{School of Physics and Astronomy, University of Southampton, Southampton, SO17 1BJ, United Kingdom}

\affiliation[b]{Particle Physics Department, Rutherford Appleton Laboratory, Chilton, Didcot, Oxon OX11 0QX, United Kingdom}

\affiliation[c]{Polydisciplinary Faculty, Laboratory of Fundamental and Applied Physics, Cadi Ayyad University, Sidi Bouzid, B.P. 4162, Safi, Morocco}

\affiliation[d]{Department of Physics and Astronomy, Uppsala University, Box 516, SE-751 20 Uppsala, Sweden}

\vspace{0.1cm}
\emailAdd{a.belyaev@soton.ac.uk}
\emailAdd{r.benbrik@uca.ac.ma}
\emailAdd{mohammed.boukidi@ced.uca.ma}
\emailAdd{mani.chakraborti@gmail.com}
\emailAdd{s.moretti@soton.ac.uk}
\emailAdd{s.semlali@soton.ac.uk}

\abstract{The recent results reported by the CMS collaboration, indicating "bumps" in the $\gamma\gamma$ and $\tau\tau$ channels at $m_\phi\approx 95$ GeV, provide interesting hints for new physics. We find that the lightest Higgs state of the general 2HDM (2HDM Type-III) can perfectly and simultaneously accommodate the two excesses alongside with the LEP long-standing anomaly observed in the $b\bar{b}$ channel while meeting all theoretical and experimental requirements. Furthermore, the study predicts an enhanced production process for the SM-like Higgs in $pp\to t\bar t H_{\rm SM}$, offering a testable hypothesis for future experiments.}
	
\FullConference{The Eleventh Annual Conference on Large Hadron Collider Physics (LHCP2023)\\
 22-26 May 2023\\
 Belgrade, Serbia\\}

\begin{document}
	\maketitle
\section{Introduction}
Following the Higgs discovery at the Large Hadron Collider (LHC), a decade-long effort has been dedicated to precisely measuring its properties. Although most observations align with the Standard Model (SM) expectations, many theories motivated the search for additional Higgs states.\\
The CMS collaboration revealed an excess with a local (global) significance of $2.9\sigma$ ($1.3\sigma$) in the di-photon final state around 95 GeV, using all LHC data collected at $\sqrt{s}=13~\text{TeV}$~\cite{CMS:2023yay}. The ATLAS group also reported a similar excess with a local significance of $1.7\sigma$~\cite{ATLAS:2018xad}. Further anomalies in the $\tau\tau$~\cite{CMS:2022goy} and $b\bar{b}$~\cite{LEPWorkingGroupforHiggsbosonsearches:2003ing} final states have been reported, respectively, by the CMS and LEP collaborations, sparking interest in explaining these within a variety of beyond SM (BSM). The 2HDM, including Type-III configuration, emerges as a promising candidate. Herein, we demonstrate the ability of the 2HDM Type-III to explain the three excesses, observed in the $\gamma\gamma$, $\tau\tau$ and $b\bar{b}$ final states. We also investigate the implication of such configuration on the Higgs production rate, particularly on $gg,qq \to H_{125}t\overline{t}$.

\section{2HDM Type-III}
The most general $SU(2)_L\times U(1)_Y$ invariant scalar potential is written in a generic basis as follows:
\vspace*{0.1cm}
\begin{eqnarray}
V &=& m_{11}^2 \Phi_1^\dagger\Phi_1+m_{22}^2\Phi_2^\dagger\Phi_2-[m_{12}^2\Phi_1^\dagger\Phi_2+{\rm h.c.}] 
+ \frac{\lambda_1}{2}(\Phi_1^\dagger\Phi_1)^2
+\frac{\lambda_2}{2}(\Phi_2^\dagger\Phi_2)^2 \nonumber \\
&+&\lambda_3(\Phi_1^\dagger\Phi_1)(\Phi_2^\dagger\Phi_2)
+\lambda_4(\Phi_1^\dagger\Phi_2)(\Phi_2^\dagger\Phi_1) 
+\left\{\frac{\lambda_5}{2}(\Phi_1^\dagger\Phi_2)^2
+\big[\lambda_6(\Phi_1^\dagger\Phi_1)
+\lambda_7(\Phi_2^\dagger\Phi_2)\big]
\Phi_1^\dagger\Phi_2+{\rm h.c.}\right\}\,. \label{pot1} \nonumber
\end{eqnarray}
Following the hermiticity of the above scalar potential and assuming CP-conservation in the 2HDM, all parameters are real-valued. After the Electroweak Symmetry Breaking (EWSB), the scalar sector involves five physical Higgs states: two CP-even states ($h$ and $H$), one CP-odd state ($A$) and a pair of charged Higgs ($H^\pm$). One can then describe the Higgs sector of the 2HDM in the physical basis by 7 parameters:
 \begin{eqnarray}
m_h, \quad m_H, \quad m_A, \quad \sin(\beta-\alpha), \quad \tan \beta = v_2/v_1\quad \text{and} \quad m_{12}^2.
 \end{eqnarray} 
$v_{1,2}$ are the Vacuum Expectation Values (VEVs) and $\alpha$ is the mixing angle of the CP-even Higgs sector.

The general structure of the Yukawa Lagrangian when both Higgs fields couple to all fermions is given by :
\begin{eqnarray}
-{\cal L}_Y &=& \bar Q_L Y^u_1 U_R \tilde \Phi_1 + \bar Q_L Y^{u}_2 U_R
\tilde \Phi_2  + \bar Q_L Y^d_1 D_R \Phi_1 
+ \bar Q_L Y^{d}_2 D_R \Phi_2 \nonumber \\
&+&  \bar L Y^\ell_1 \ell_R \Phi_1 + \bar L Y^{\ell}_2 \ell_R \Phi_2 + {\rm H.c.},
\label{eq:Yu}
\end{eqnarray}

The presence of an extra doublet within the Yukawa sector gives rise to unwanted tree level Higgs mediated Flavour Changing Neutral Currents (FCNCs), which are strongly constrained by various experiments. In order to keep the FCNCs under control, we adopt the Cheng-Sher ansatz~\cite{Cheng:1987rs,Diaz-Cruz:2004wsi}, which assumes a specific form of the non-diagonal Yukawa terms, $g_{ij} \propto \sqrt{m_i m_j}/v \chi_{ij}^f$. 

Different public tools were used to check all theoretical and experimental constraints restricting the 2HDM parameter space:
\begin{itemize}
	\item \texttt{2HDMC}~\cite{Eriksson:2009ws} enables to test Unitarity, Perturbativity, Vaccuum Stability and EW Precision Observables,
	\vspace{-0.2cm} 
	\item \texttt{HiggsTools}~\cite{Bahl:2022igd} provides a new framework to call both \texttt{HiggsBounds-6}  and \texttt{HiggsSignals-3}  to apply constraints stemming from direct Higgs searches at (hadron and lepton) colliders, and from Higgs boson current signal strength measurements, respectively,
	\vspace{-0.2cm}
	\item \texttt{SuperIso}\cite{Mahmoudi:2008tp} to account for constraints from various flavour physics observables.
\end{itemize}

\section{Numerical results}

To investigate whether the 2HDM Type-III can perfectly explain the three excesses observed in the $\gamma\gamma$, $\tau\tau$ and $b\bar{b}$ final states at once, we perform a $\chi^2$ analysis for each channel, 
\begin{eqnarray}
\chi^2_{\gamma\gamma,\tau\tau,b\bar{b}}=\left(\mu_{\gamma\gamma,\tau\tau,b\bar{b}}-\mu_{\gamma\gamma,\tau\tau,b\bar{b}}^\mathrm{ exp}\right)^2/\left(\Delta\mu^\mathrm{exp}_{\gamma\gamma,\tau\tau,b\bar{b}}\right)^2.
\end{eqnarray}
Based on the resulting $\chi^2_{\gamma\gamma+\tau\tau+b\bar{b}}= \chi^2_{\gamma\gamma}+\chi^2_{\tau\tau}+\chi^2_{b\bar{b}} $, we will conclude if the model can simultaneously describe the three excesses. The signal strength $\mu_{XX}$\footnote{Assuming the Narrow Width Approximation (NWA), the signal strength $(\mu_{XX})$ can be expressed in terms of cross section ($\sigma$) and branching ratio ($BR$), i.e.,  $\mu_{XX} = \sigma \times BR$.} with $XX=\gamma\gamma,~\tau\tau,~b\bar{b}$ is written as follows:
\begin{eqnarray}
\mu_{{b\bar{b}}}&=&\frac{\sigma_{\rm 2HDM}(e^+e^-\to Zh )}{\sigma_{\rm SM}(e^+e^-\to Zh_{SM})}\times \frac{{\cal BR}_{\rm 2HDM}(h \to b\bar{b})}{{\cal BR}_{\rm SM}(h_{\rm SM}\to b\bar{b})} =\left|c_{h ZZ}\right|^2\times \frac{{\cal BR}_{\rm 2HDM}(h \to b\bar{b})}{{\cal BR}_{\rm SM}(h_{\rm SM}\to b\bar{b})},\label{mu_lep}\\\nonumber\\
\mu_{\mathrm{\tau\tau}}&=&\frac{\sigma_{\rm 2HDM}(gg\to h )}{\sigma_{\rm SM}(gg\to h_{\rm SM})}\times \frac{{\cal BR}_{\rm 2HDM}(h \to \tau\tau)}{{\cal BR}_{\rm SM}(h_{\rm SM}\to \tau\tau)} =\left|c_{h tt}\right|^2\times \frac{{\cal BR}_{\rm 2HDM}(h \to \tau\tau)}{{\cal BR}_{\rm SM}(h_{\rm SM}\to \tau\tau)},\\\nonumber\\
\mu_{\mathrm{\gamma\gamma}}&=&\frac{\sigma_{\rm 2HDM}(gg\to h )}{\sigma_{\rm SM}(gg\to h_{\rm SM})}\times \frac{{\cal BR}_{\rm 2HDM}(h \to \gamma\gamma)}{{\cal BR}_{\rm SM}(h_{\rm SM}\to \gamma\gamma)} =\left|c_{h tt}\right|^2\times \frac{{\cal BR}_{\rm 2HDM}(h \to \gamma\gamma)}{{\cal BR}_{\rm SM}(h_{\rm SM}\to \gamma\gamma)},\label{mu_cms}
\end{eqnarray}
$h_{SM}$ represents the SM Higgs-boson. Its mass  is rescaled to that of the lightest Higgs boson $h$ of the 2HDM Type-III. The experimental measurements of the signal strengths are:
\begin{eqnarray}
\mu_{\gamma\gamma}^{\mathrm{exp}}=\mu_{\gamma\gamma}^{\mathrm{ATLAS+CMS}} = 0.27^{+0.10}_{-0.09},
\quad \quad\quad \mu_{\tau\tau}^{\mathrm{exp}} = 1.2 \pm 0.5, \quad \quad\quad  \mu_{b\bar{b}}^{\mathrm{exp}} = 0.117 \pm 0.057,
\end{eqnarray}

\begin{figure}[h!]
	\centering
	\includegraphics[scale=0.3]{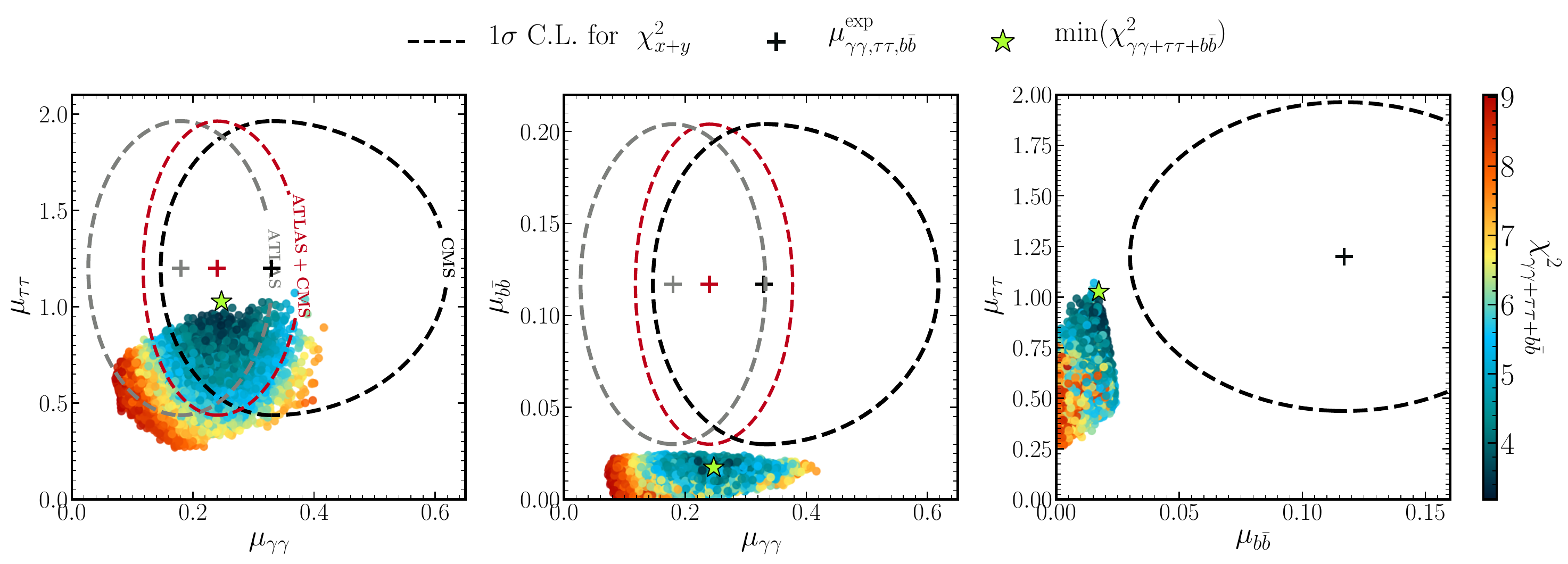}	
	\caption{
		The colour map  of $\chi^2_{\gamma\gamma+\tau\tau}$
		in the  the 
		($\mu_{\tau\tau}$-$\mu_{\gamma\gamma}$),
		plane of the 
		signal strength parameters for  2HDM Type-III parameter space under study.
		The dashed ellipses define the
		regions consistent with the excess at 
		1$\sigma$ C.L.
		The black, gray and red contours are for $\chi^2$ constructed
		using $\mu_{\gamma\gamma}^{\mathrm{CMS}}$, $\mu_{\gamma\gamma}^{\mathrm{ATLAS}}$ and $\mu_{\gamma\gamma}^{\mathrm{CMS+ATLAS}}$ signal strengths respectively. Position of  $\chi^2_{95\mathrm{min}}$  is indicated  by green star.
		\label{fig1}}
\end{figure}

Fig.~\ref{fig1} displays the parameter space\footnote{The list of the parameters ranges can be found in~\cite{Belyaev:2023xnv}} satisfying the aforementioned theoretical and experimental constrains in different planes. Clearly, the minimum of $\chi_{95}^2$, marked by a green star, alongside with other points (coloured by dark blue) fall within 1$\sigma$ confidence level (corresponding to $\chi^2<3.53$). This demonstrates the ability of the 2HDM Type-III to accommodate the three excesses simultaneously at 1$\sigma$ confidence level. 

\begin{figure}[h!]	
	\centering
	\includegraphics[scale=0.3]{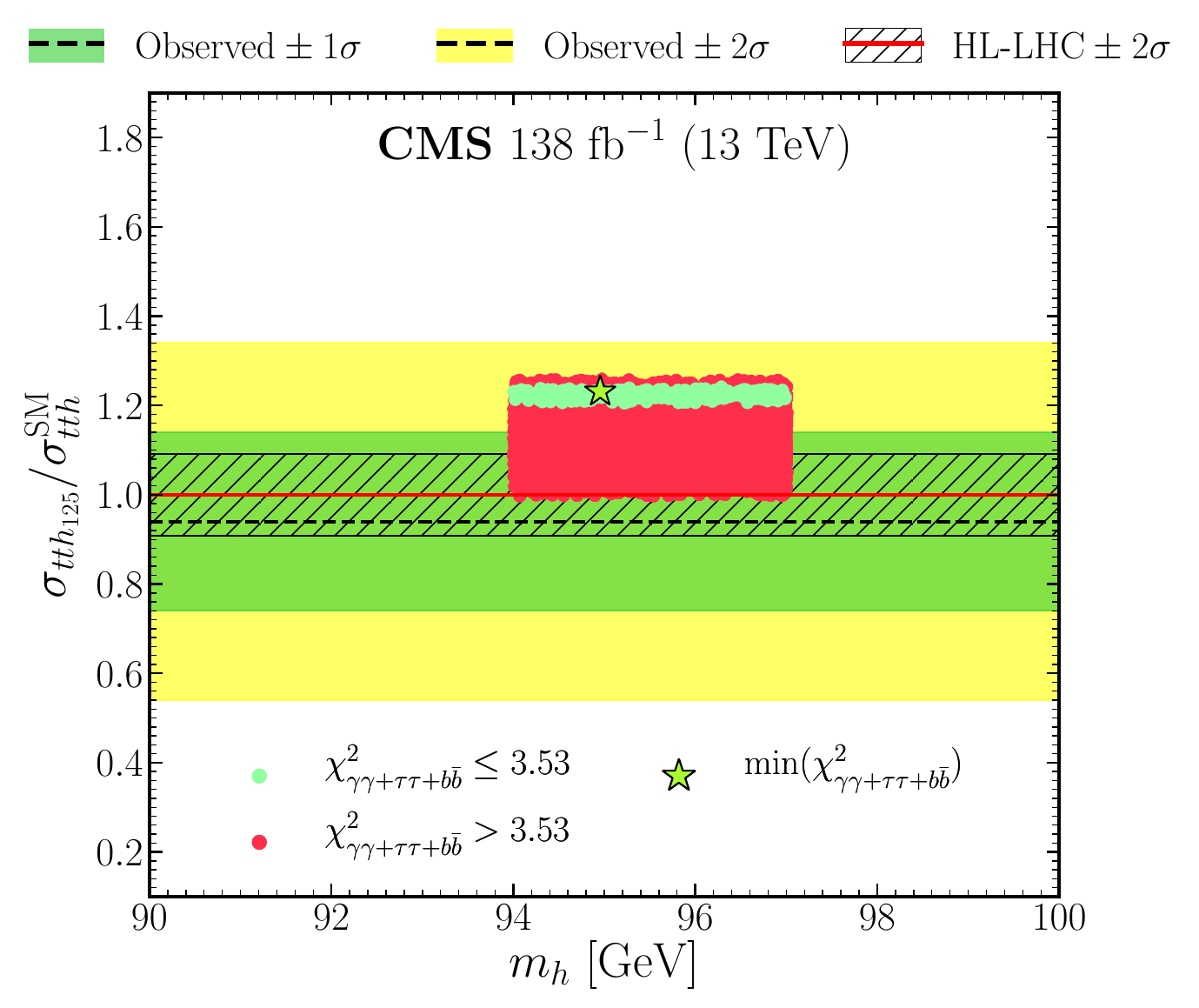}
	\caption{Predicted signal strength $\sigma_{tth_{125}}/\sigma_{tth}^{\mathrm{SM}}$ in relation to the three observed excesses. The black dashed line represents the observed value from CMS \cite{CMS:2022dwd} while the lime green (yellow) band denotes the $1\sigma$ ($2\sigma$) range around the observed value. The hatched area illustrates the 95\% probability sensitivity of the HL-LHC \cite{Cepeda:2019klc} to the normalised cross section $\sigma_{tth_{125}}/\sigma_{tth}^{\mathrm{SM}}$, centered on the value of the SM (depicted as a solid red line). .}\label{fig2} 
\end{figure}

Fig.~\ref{fig2} shows the normalised cross section ($\sigma_{tth_{125}}/\sigma_{tth}^{SM}$) as a function of $m_h$. Notably, it is intriguing to observe that the data points corresponding to the region explaining the three excesses at $1\sigma$, marked in green, lie within the $2\sigma$ range of the CMS measurement. However, these points exhibit a noticeable deviation from the $2\sigma$ level projected for the HL-LHC. This outcome is due to the enhancement in the coupling of the 125 GeV scalar to top quarks. Hence, for these points that deviate significantly  from the Standard Model (SM) prediction, the HL-LHC would be able to discern between the properties of the SM-like of the $h_{125}$ GeV  and the prediction of the 2HDM Type-III.

\clearpage

\vspace*{-0.2cm}
\section*{Conclusion}

Within the 2HDM Type-III, we have demonstrated that the excesses observed in the di-photon, di-tau and $b\bar{b}$ channels can be produced  by the lightest CP-even Higgs state, $h$, with a mass close to 95 GeV, while being in agreement with theoretical requirements and up-to-date experimental constraints. Overall, the embedded 95 GeV resonance and the predicted enhancement in the Yukawa coupling to top quarks make this scenario particularly intriguing for further experimental investigations at the LHC, HL-LHC, and future particle colliders.

\end{document}